# Saturation of the quantum Cramér-Rao bound for distributed sensing via error sensitivity in SU(1,1)-SU($m$) interferometry

Girish S. Agarwal [*]

*Institute for Quantum Science and Engineering, Department of Physics and Astronomy,*
*Department of Biological and Agricultural Engineering, Texas A&M University, College Station, Texas 77843, USA*

Breaking the standard quantum limit in the sensing of parameters at different spatial locations, such as in a quantum network, is of great importance. Using the framework of quantum Fisher information, many strategies based on squeezed quantum probes and multipath multiphoton or multiqubit entangled states have been considered. In this context there is always the question of what the simplest measurement is that would saturate the quantum Cramér-Rao bound (QCRB). The simplest quantity to measure would be characteristics of photon flux or population distribution in the case of qubits. Previous studies have shown that the error sensitivity in SU(1,1) interferometry, also known by several other terms such as nonlinear interferometry and time-reversed measurements, does saturate the QCRB for single parameters such as phase, displacement, and loss. In this work we reveal the great utility of generalized SU(1,1) interferometry in distributed sensing. Generalized SU(1,1) interferometry is a combination of SU($m$) and SU(1,1) elements, where $m$ is the number of nodes in the network. The SU($m$) element is used to produce distributed entanglement starting from a squeezed photonic or matter probe. We demonstrate how error sensitivity measurement or more precisely the method of moment sensitivity at just one output port can saturate or nearly saturate the QCRB and thus results in Heisenberg sensitivity of network sensing.

## I. INTRODUCTION

Use of quantum probes in estimating the sensitivity of parameters, especially phases and rotations, has been extremely successful. The phase can be in an interferometric setup [1–4], which could be a photonic or an atomic interferometer [5]. The squeezed states are very special here and one can have in principle optimum quantum advantage, i.e., quantum sensing can reach Heisenberg-limited sensitivity. Estimation of many parameters is more challenging [6–9]. Important examples are rotations in three dimensions [10,11] and forces, as force is a vectorial property. It is also not clear what quantum probe would be best for simultaneous measurement of several parameters. An even more complex problem is the sensing of parameters of a quantum network [12–16], which is of fundamental importance in many applications such as the construction of a quantum-enhanced world clock [17] and beam tracking [18]. Two important classes of networks have been investigated [19–24]. These include networks where the key quantity at a node of the network is a phase and the displacement related to some unknown fields [19–27]. The ultimate goal is to reach the Heisenberg limit. Quantum states of light, in particular, both entangled and squeezed states, spin squeezed states, and multiqubit entangled states, have been found useful.

A quantum meteorological framework based on quantum Fisher information and the quantum Craméer-Rao bound (QCRB) has been very successful in sensitivity estimates of the unknown parameters [28–33]. This enables us to find appropriate quantum probes leading to estimates well beyond the standard quantum limit. Having found the best quantum probe, we need to search for the experimental scheme in which a simple measurement of the error sensitivity would saturate the QCRB. Our emphasis is on error sensitivity, or more precisely the method of moments sensitivity derived from the measurement of photon flux and fluctuations in it. For brevity we will continue to use the term error sensitivity. Error sensitivity is the preferred quantity as flux is easily measured without needing, for example, number-resolving detectors. Flux measurements are simpler than, say, homodyne measurements. The resource requirement for the SU(1,1)-SU($m$) scheme is less stringent than in homodyne measurements as flux measurement is much simpler than quadrature measurement. Besides, a single flux measurement is adequate, whereas the homodyne scheme requires measurement at all ports, though both schemes use beam splitters and optical parametric amplifiers (OPAs). Recent works have found answers to this for estimates of a single parameter-like phase and displacement using SU(1,1) interferometers [3,26,34–47]. A study of this is warranted for multiple parameters or for estimating spatially distributed parameters as in a network. We report such a study in this paper.

[*]Contact author: girish.agarwal@ag.tamu.edu

$$|0\rangle \xrightarrow{\hat{S}(\zeta)} |\psi_\zeta\rangle \xrightarrow{\hat{D}(\alpha) \text{ or } \hat{R}(\phi)} |\Phi(\zeta,\eta)\rangle \xrightarrow{\hat{S}^{-1}(\zeta)} |\psi(\zeta,\eta)\rangle.$$

FIG. 1. Schematics of the SU(1,1) interferometer.

The organization of this paper is as follows. We start in Sec. II by introducing how SU(1,1) interferometry has been successful in Heisenberg-limited sensing of single parameters like phase and displacement. We use single-mode squeezed vacuum [48] as the resource quantum probe. In Sec. III we show that the error sensitivity measurement of photon flux at a single output port of the SU(1,1)-SU($m$) interferometer yields Heisenberg-limited sensitivity, i.e., saturation of the QCRB, of the average force fields in a quantum network. In Sec. IV we consider distributed sensing of the average phase using an SU(1,1)-SU($m$) interferometer and show that the error sensitivity measurement of photon fluxes gives nearly the Heisenberg-limited sensitivity. One does not need number resolving or homodyne measurements in SU(1,1)-SU($m$) interferometry with squeezed vacuum as quantum probes. Our scheme should also be applicable to distributed sensing of angular displacements of beams using quantum states of orbital angular momentum [49]. In the Appendixes we give simplified and easily followed calculations of the quantum Fisher information for the two networks considered in this paper.

## II. SATURATION OF THE CRAMÉR-RAO BOUND VIA ERROR SENSITIVITY MEASUREMENT IN NONLINEAR SU(1,1) INTERFEROMETRY: SIMPLE EXAMPLES

In this section we consider a scheme which can saturate the QCRB via very simple measurements of the output signal and fluctuation in the output signal. The signal for our purpose will be the intensity of the output signal. These measurements do not require photon-number-resolving detectors and are much simpler to implement than homodyne measurements. The scheme is motivated by the well-known time-reversed metrology scheme used extensively for the estimate of a single parameter such as the displacement $\alpha = |\alpha|e^{i\theta}$ resulting from electric fields and phase $\phi$ corresponding to rotations. This scheme is shown in Fig. 1.

In Fig. 1 the parameter $\eta$ represents the unknown parameter whose sensitivity is being studied by using the quantum probe, which is a single-mode squeezed state $|\psi_\zeta\rangle$. The squeezing operator is given by $S(\zeta) = \exp(\frac{1}{2}\zeta a^{\dagger 2} - \frac{1}{2}\zeta^* a^2)$, with $\zeta = re^{i\beta}$. Here $r$ is the squeezing parameter. Note that the squeezing operations are elements of the SU(1,1) group and are implemented by opitcal parametric amplifiers. The parameter $\eta = \alpha$ for displacement generated by the unitary operator $\hat{D}(\alpha) = \exp(\alpha\hat{a}^\dagger - \alpha^*\hat{a})$ or $\eta = \phi$ for rotation $\hat{R}(\phi) = \exp(-i\phi\hat{a}^\dagger\hat{a})$ or for phase estimation. In the time-reversed scheme, we add the additional step represented by the inverse of the squeezing $\hat{S}^{-1}(\zeta)$, which enables us to achieve Heisenberg-limited sensitivity via the measurement of the signal intensity and its fluctuations.

We would like to clarify that in the classic paper by Yurke *et al.* [34] a two-mode representation of the SU(1,1) group is used, whereas we work with a single-mode squeezed vacuum. This is because the SU(1,1) group has realizations in terms of both a single bosonic oscillator, related to a single-mode squeezed vacuum, and two oscillators, related to a two-mode squeezed vacuum. The Lie algebra for SU(1,1) is given by

$$[K_-, K_+] = 2K_3, \quad [K_3, K_+] = K_+, \quad [K_3, K_-] = -K_-. \tag{1}$$

The $K$ operators can be written in terms of single-mode boson algebra $-K_- = \frac{aa}{2}$, $K_+ = \frac{a^\dagger a^\dagger}{2}$, and $K_3 = \frac{aa^\dagger + a^\dagger a}{2}$. Hence the single-mode squeezing operator, given by $S(\xi) = \exp(\frac{1}{2}\xi a^{\dagger 2} - \frac{1}{2}\xi^* a^2)$, with $\xi = re^{i\varphi}$, corresponds to the SU(1,1) group. The two-mode realization of SU(1,1) is discussed in, for example [48].

If the parameter to be estimated is displacement $\alpha = |\alpha|e^{i\theta}$, then a calculation shows that the final state is a coherent state

$$|\psi(\zeta,\alpha)\rangle = \langle |\alpha| e^{i\theta}\big(\cosh r - \sinh r\, e^{i\beta} e^{-2i\theta}\big)\rangle,$$

with an amplitude that depends on the parameter to be estimated, the squeezing parameter $r$, and its direction of squeezing given by $\beta$. In particular, if $\beta - 2\theta = \pi$, then

$$\psi(\zeta,\alpha)\rangle = ||\alpha| e^{i\theta} e^r\rangle. \tag{2}$$

Thus the estimation of amplitude $|\alpha|$ is amplified by a factor $e^r$. The error sensitivity $\Delta|\alpha|$ is defined in terms of the signal $S = \langle\hat{a}^\dagger\hat{a}\rangle$ and the fluctuation in the signal $(\Delta S)^2 = \langle(\hat{a}^\dagger\hat{a})^2\rangle - \langle\hat{a}^\dagger\hat{a}\rangle^2$, i.e.,

$$\Delta|\alpha| = \Delta S \Big/ \left(\frac{\partial S}{\partial|\alpha|}\right). \tag{3}$$

Since (2) is a coherent state, the error sensitivity becomes $S = |\alpha|^2 e^{2r}$, $(\Delta S)^2 = S = |\alpha|^2 e^{2r}$, and hence

$$\Delta|\alpha| = \tfrac{1}{2}e^{-r}. \tag{4}$$

We have used the property that the photon-number fluctuation in the coherent state is equal to the photon number. The foregoing assumes fixed $\theta$ and that the direction of squeezing of the probe has to be chosen depending on the angle $\theta$. We next calculate the quantum Fisher information using the unitary evolution corresponding to displacement. We write $\hat{D}(\alpha)$ as

$$\hat{D}(\alpha) = \exp[|\alpha|(\hat{a}^\dagger e^{i\theta} - \hat{a}e^{-i\theta})] = e^{i|\alpha|\hat{G}},$$

where $\hat{G} = -i(\hat{a}^\dagger e^{i\theta} - \hat{a}e^{-i\theta})$. The quantum Fisher information (QFI), defined by

$$F_Q = 4\langle\psi_\zeta|(\hat{G} - \langle\hat{G}\rangle)^2|\psi_\zeta\rangle, \tag{5}$$

can be calculated using the properties of the single-mode squeezed vacuum state $|\psi_\zeta\rangle = S(\zeta)|0\rangle$,

$$\langle\hat{G}\rangle = -i\langle\psi_\zeta|\hat{a}^\dagger e^{i\theta} - \hat{a}e^{-i\theta}|\psi_\zeta\rangle = 0.$$

A simple calculation shows that if $\beta - 2\theta = \pi$,

$$F_Q = 4e^{2r}, \tag{6}$$

and hence

$$(\Delta|\alpha|)_{\text{QCRB}} \geqslant \frac{1}{\sqrt{4e^{2r}}} = \frac{1}{2}e^{-r}. \tag{7}$$

On comparing (7) and (3) we see that the SU(1,1) interferometer measurement of error sensitivity saturates the QCRB.

The bound (7) or (3) is also the Heisenberg bound on the sensitivity of the displacement measurement.

For completeness, we repeat similar arguments for the sensitivity of the phase estimator. In this case the QFI is obtained by noting that $\hat{G} = \hat{a}^\dagger \hat{a}$ and hence

$$F_Q = 4\langle\psi_\zeta|(\hat{a}^\dagger\hat{a})^2|\psi_\zeta\rangle - 4\langle\psi_\zeta|(\hat{a}^\dagger\hat{a})|\psi_\zeta\rangle^2. \tag{8}$$

The QFI is proportional to the number fluctuation in the squeezed vacuum $|\psi_\zeta\rangle$. These are known [48] and hence

$$F_Q = 8\sinh^2(r)\cosh^2(r),$$
$$(\Delta\phi)_{\text{QCRB}} \geqslant \frac{1}{2\sqrt{2}\sinh(r)\cosh(r)}. \tag{9}$$

For error sensitivity we need to find the signal $\langle\hat{a}^\dagger\hat{a}\rangle$ and its fluctuations $\Delta S$ in the state $|\psi(\zeta,\eta)\rangle$. Note that we can transform the expectation values in the final state $|\psi(\zeta,\eta)\rangle$ to the values in the initial vacuum state as follows:

$$\langle\psi(\zeta,\eta)|\hat{G}|\psi(\zeta,\eta)\rangle = \langle 0|\hat{\hat{G}}|0\rangle, \tag{10}$$

$$\hat{\hat{G}} = \hat{S}^\dagger(\zeta)e^{i\phi\hat{a}^\dagger\hat{a}}\hat{S}(\zeta)(\hat{G})\hat{S}^\dagger(\zeta)e^{-i\phi\hat{a}^\dagger\hat{a}}\hat{S}(\zeta). \tag{11}$$

In particular, the annihilation operator $\hat{a}$ is transformed to

$$\hat{\hat{a}} = A\hat{a}^\dagger + B\hat{a},$$
$$A = e^{i\chi}\cosh r\sinh r(e^{-i\phi} - e^{i\phi}), \tag{12}$$
$$B = e^{-i\phi}\cosh^2 r - e^{i\phi}\sinh^2 r.$$

Using (11), we can obtain the signal and fluctuations as

$$S = |A|^2 = 4\sin^2\phi\cosh^2 r\sinh^2 r,$$
$$\Delta S = |A|\sqrt{1 + |A|^2 + |B|^2}. \tag{13}$$

The error sensitivity in the limit $\phi \to 0$ is found to be

$$\Delta\phi = \frac{1}{2\sqrt{2}\cosh r\sinh r}, \tag{14}$$

which coincides with the QCRB (9).

The two examples show that the single-mode squeezed vacuum is the optimum probe state for estimating both the displacement and phase using a nonlinear SU(1,1) interferometer. The measured error sensitivity bound coincides with the QCRB. Thus the nonlinear SU(1,1) interferometer is ideal in the sense that its error sensitivity bound obtained for the measurements of the mean intensity and its fluctuations saturates the QCRB. This nonlinear interferometer avoids the need for photon-number quantum probes, any photon-number resolving, or homodyne measurements. The discussion presented here would also apply to other parameters such as angular displacements [49] or beam tracking [18]. This requires generalizing the present discussion to a two-mode squeezed vacuum and use of spatial light modulator to the generated probe field. A simpler situation is discussed in [49].

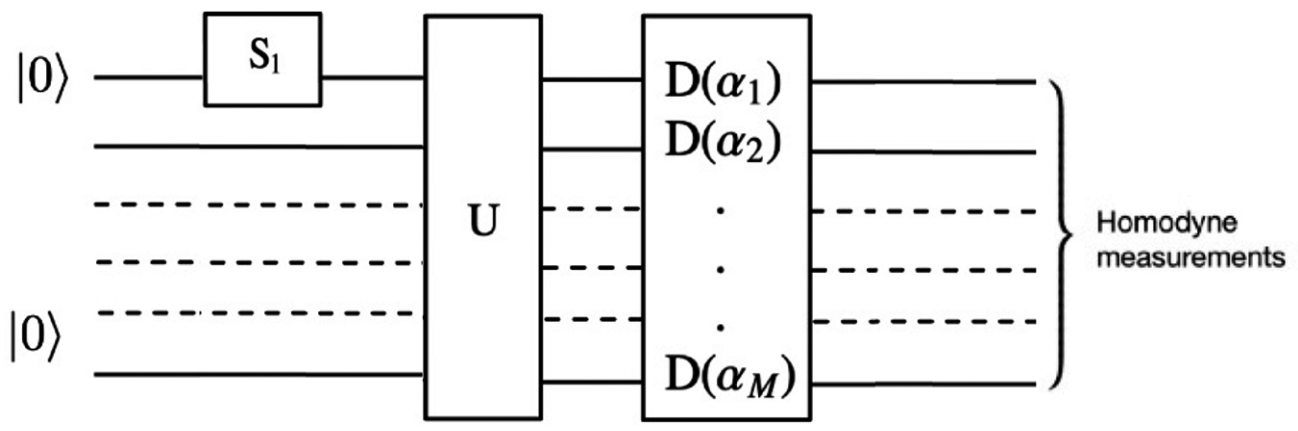

FIG. 2. Traditional measurements of the network average with all operations described in the text.

## III. DISTRIBUTED SENSING OF FORCE FIELDS OR DISPLACEMENT PARAMETERS: SATURATION OF QCRB VIA ERROR SENSITIVITY MEASUREMENTS IN THE SU(1,1)-SU(*m*) INTERFEROMETER

We now generalize the simple analysis presented in the preceding section to distributed sensing of displacement parameters. We sketch in Figs. 2 and 3 the standard measurement scheme and our proposed nonlinear interferometer scheme.

In this scheme $\hat{U}$ is an unitary operator which distributes a single-mode quantum probe over different modes described by the annihilation operators $\hat{a}_1, \hat{a}_2, \ldots, \hat{a}_m$ and thus different modes get entangled. Such entangled modes probe the network of displacements. Clearly $U$ is a unitary matrix in $m$-dimensional space, i.e., it has dimensions $m \times m$. Thus the transformation matrix $U$ belongs to the group SU($m$). Each mode of the network is being probed by one mode. Note that the unitary operator $U$ is an element of the group SU($m$). For simplicity of analysis we will assume that the $\alpha$ are real and the angle $\beta = \pi$. This is the choice we made in Sec. II for single displacement estimation [cf. Eq. (2)]. Our goal is to saturate the QCRB for $\bar{\alpha}$ via error sensitivity measurement on mode 1, as indicated in Fig. 3. We would choose the unitary matrix such that

$$U_{j1} = \frac{1}{\sqrt{M}}. \tag{15}$$

This choice distributes the input photon number equally in all modes as

$$\hat{b}_i = \sum_l \hat{U}_{il}\hat{a}_l,$$
$$\langle\hat{b}_i^\dagger\hat{b}_i\rangle = \sum_l\sum_{l'}\hat{U}_{il}^*\hat{U}_{il'}\langle\hat{a}_l^\dagger\hat{a}_{l'}\rangle$$
$$= \sum_l\sum_{l'}\hat{U}_{il}^*\hat{U}_{il'}\delta_{ll'}\langle\hat{a}_l^\dagger\hat{a}_l\rangle. \tag{16}$$

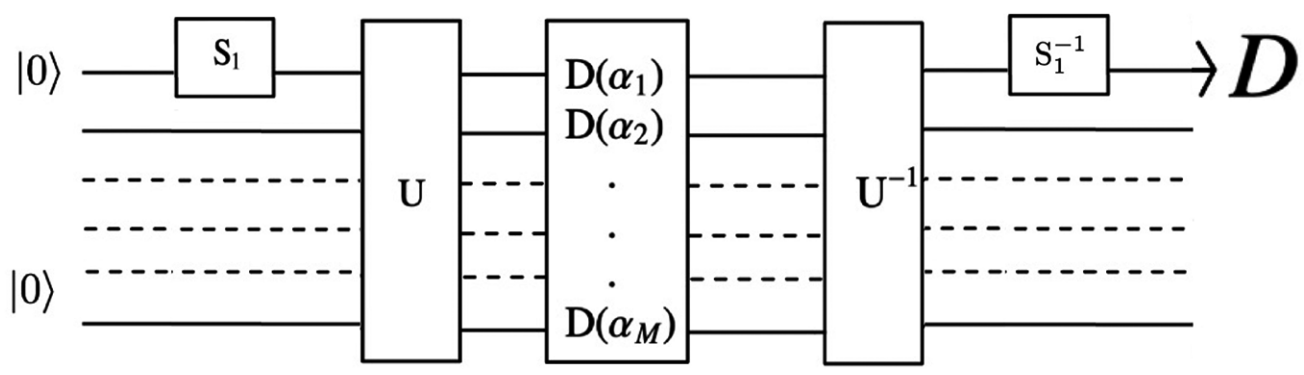

FIG. 3. SU(1,1)-SU($m$) interferometric measurement of the error sensitivity from the signal at port 1.

Since only the mode $\hat{a}_1$ is occupied, Eq. (16) yields

$$\langle \hat{b}_i^\dagger \hat{b}_i \rangle = \hat{U}_{i1}^* \hat{U}_{i1} \langle \hat{a}_1^\dagger \hat{a}_1 \rangle = \frac{1}{M} \langle \hat{a}_1^\dagger \hat{a}_1 \rangle. \tag{17}$$

Here we study the most sensitive measurement of the average of the displacement over the network, i.e., the quantity $\bar{\alpha} = \sum_j \frac{\alpha_j}{M}$. We proceed by calculating the output state of the multimode system by following the series of transformations as shown in Fig. 3. Let $|\{0\}\rangle$ represent the vacuum state of all the modes. Then the output state will be

$$|\psi\rangle = \hat{S}_1^{-1}(\zeta)\hat{U}^{-1} \prod_j D_j(\alpha_j)\hat{U}\hat{S}_1(\zeta)|\{0\}\rangle. \tag{18}$$

Here $\hat{S}_1(\zeta)$ is the single-mode squeezing operator of the mode 1. Note that $\prod_j D_j(\alpha_j) = \exp[\sum_j(\alpha_j \hat{a}_j^\dagger - \alpha_j \hat{a}_j)]$. The unitary transformation in (18) will transform $\hat{a}_j$ into new operators $\hat{\tilde{a}}_j$,

$$\hat{\tilde{a}}_j = \hat{S}_1^{-1}(\zeta)\hat{U}^{-1}\hat{a}_j \hat{U}\hat{S}_1(\zeta), \tag{19}$$

and hence

$$|\psi(\zeta)\rangle = \exp\left(\sum_j (\alpha_j \hat{\tilde{a}}_j^\dagger - \alpha_j \hat{\tilde{a}}_j)\right)|\{0\}\rangle. \tag{20}$$

We can write $\hat{\tilde{a}}$ as

$$\begin{aligned}
\hat{\tilde{a}}_j &= \sum_l U_{jl}\hat{S}_1^{-1}(\zeta)\hat{a}_l \hat{S}_1(\zeta) \\
&= \sum_{l\neq 1} U_{jl}\hat{a}_l + U_{j1}\hat{S}_1^{-1}(\zeta)\hat{a}_1\hat{S}_1(\zeta) \\
&= \sum_{l\neq 1} U_{jl}\hat{a}_l + U_{j1}(\hat{a}_1 \cosh r - \hat{a}_1^\dagger \sinh r),
\end{aligned}$$

and hence

$$\begin{aligned}
\sum_j \alpha_j \hat{\tilde{a}}_j &= \sum_j \sum_{l\neq 1} \hat{a}_l U_{jl}\alpha_j + \sum_j \alpha_j U_{j1}(\hat{a}_1 \cosh r - \hat{a}_1^\dagger \sinh r) \\
&= \bar{\alpha}\sqrt{M}(\hat{a}_1 \cosh r - \hat{a}_1^\dagger \sinh r) + \sum_{l\neq 1} \hat{a}_l \sum_j U_{jl}\alpha_j.
\end{aligned} \tag{21}$$

On substituting (21) in (20) and rearranging, we find that the state $|\psi\rangle$ is a product of coherent states $|\xi_j\rangle$ with

$$\xi_1 = \bar{\alpha}\sqrt{M}e^r, \quad \xi_l = \sum_j U_{jl}\alpha_j, \ l \neq 1. \tag{22}$$

Note that states of the modes with $l \neq 1$ do not depend on the squeezing parameter, whereas the state of the initially squeezed mode becomes a coherent state $|\xi_1\rangle$ with a magnification factor that depends on the network average $\bar{\alpha}$ of the displacement parameters and $\sqrt{M}e^r$, where $M$ is the number of modes and $r$ is determined by the squeezing of the input single mode. The error sensitivity of the output mode 1 can be calculated from the properties of the coherent state $\langle\beta|$. For any coherent state the mean number of photons is the signal $S = \langle a^\dagger a\rangle = |\beta|^2$ and fluctuation in photon number $\Delta S = \{\langle (a^\dagger a)^2\rangle - [\langle (a^\dagger a)\rangle]^2\}^{1/2} = \sqrt{\langle a^\dagger a\rangle}$. If $\beta$ is a function of the parameter $\epsilon$ to be estimated, then error sensitivity is defined as $\Delta\epsilon = \Delta S / \frac{\partial S}{\partial \epsilon}$. In the present case $\beta = \xi_1 = \bar{\alpha}\sqrt{M}e^r$, $\epsilon = \bar{\alpha}$, and hence $|\beta|^2 = \bar{\alpha}^2 M e^{2r}$. Therefore, $\frac{\partial S}{\partial \bar{\alpha}} = 2\bar{\alpha}Me^{2r}$ and $\Delta S = \bar{\alpha}\sqrt{M}e^r$, leading to

$$\Delta\bar{\alpha} = \frac{1}{2\sqrt{M}e^r}. \tag{23}$$

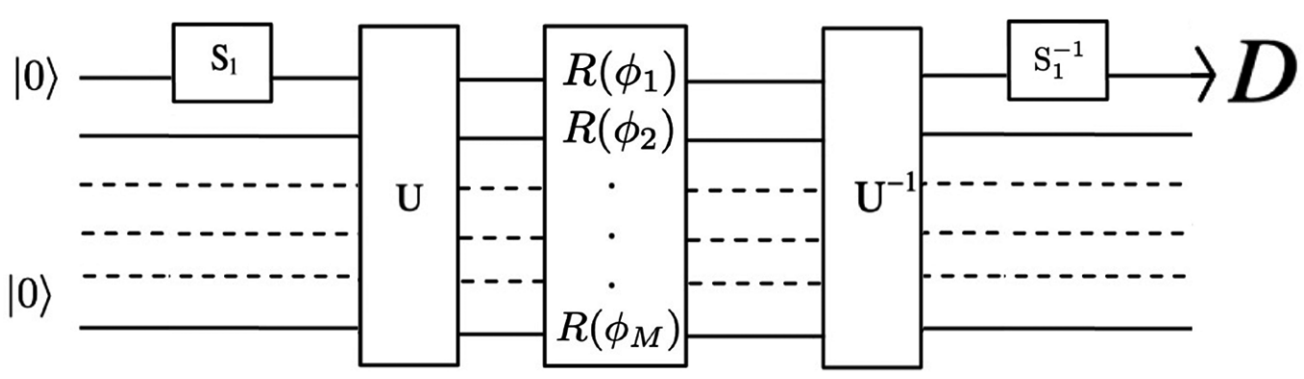

FIG. 4. Sensing of the average phase in a network via measurement of the error sensitivity in the SU(1,1)-SU($m$) interferometer.

The error sensitivity (23) is identical to the Heisenberg-limited sensitivity (A10) of the network average $\bar{\alpha}$. The details of the QCRB for $\bar{\alpha}$ are given in Appendix A. It is remarkable that for the measurement scheme shown in Fig. 3 we find that the network displacement average error sensitivity bound is equal to the QCRB, which in turn is equal to Heisenberg-limited sensitivity.

The SU($m$) is implemented by using beam splitters, as shown in experiment in [19,20]. The SU[1,1] is implemented by using OPAs. This is quite standard now as many experiments [42–44] have implemented the SU(1,1) interferometer. Further, in many experiments one uses a configuration so that one does not need two OPAs [43,50]. Thus, everything about the SU(1,1)- SU($m$) interferometer is within the realm of current technology.

Finally, we note the result for error sensitivity if a homodyne measurement is done at port 1. The signal $S_{\rm HD}$ would be

$$S_{\rm HD} = \left\langle \frac{a^\dagger + a}{\sqrt{2}} \right\rangle = \sqrt{2}\bar{\alpha}\sqrt{M}e^r, \tag{24}$$

where the average is with respect to the coherent state $\bar{\alpha}\sqrt{M}e^r$. The fluctuation in the signal will be

$$(\Delta S_{\rm HD})^2 = \left\langle \left(\frac{a^\dagger + a}{\sqrt{2}}\right)^2 \right\rangle - \left\langle \left(\frac{a^\dagger + a}{\sqrt{2}}\right) \right\rangle^2 = \frac{1}{2} \tag{25}$$

and hence

$$(\Delta\bar{\alpha})_{\rm HD} = \Delta S_{\rm HD} \Big/ \frac{\partial S_{\rm HD}}{\partial \bar{\alpha}} = \frac{1}{2\sqrt{M}e^r}, \tag{26}$$

which coincides with (23) obtained from photon flux measurements.

## IV. SATURATION OF THE CRAMÉR-RAO BOUND FOR THE AVERAGE PHASE OF THE NETWORK BY THE SU(1,1)-SU($m$) INTERFEROMETRIC ERROR SENSITIVITY ESTIMATE

In this section we show how the quantum Cramér-Rao bound (B14) can be saturated via the error sensitivity measurement using nonlinear interferometry. The nonlinear interferometric setup is shown in Fig. 4.

We choose $U$ in such a way that each mode after $U$ has the same number of averaged photons, i.e., $U_{j1} = \frac{1}{\sqrt{M}}$. The series of operations shown in Fig. 4 will give the wave function for all the output modes as

$$|\psi\rangle = \hat{S}_1^{-1}\hat{U}^{-1}\prod_j \hat{R}(\phi_j)\hat{U}\hat{S}_1|\{0\}\rangle. \tag{27}$$

Therefore, the expectation value of any operator $\hat{G}$ will be

$$\begin{aligned}\langle\psi|\hat{G}|\psi\rangle &= \langle\{0\}|\hat{\hat{G}}|\{0\}\rangle,\\ \hat{\hat{G}} &= \hat{S}_1^\dagger\hat{U}^\dagger\prod_j \hat{R}^\dagger(\phi_j)(\hat{U}^{-1})^\dagger(\hat{S}_1^{-1})^\dagger\hat{G}\hat{S}_1^{-1}\hat{U}^{-1}\\ &\quad\times\prod_j \hat{R}(\phi_j)\hat{U}\hat{S}_1\\ &= \hat{S}_1^\dagger\hat{U}^\dagger\prod_j \hat{R}^\dagger(\phi_j)\hat{U}\hat{S}_1\hat{G}\hat{S}_1^{-1}\hat{U}^{-1}\prod_j \hat{R}(\phi_j)\hat{U}\hat{S}_1.\end{aligned} \tag{28}$$

We define the signal $S$ as the photon number at the detector $D$, i.e., $\langle\psi|\hat{a}_1^\dagger\hat{a}_1|\psi\rangle$. Thus the error sensitivity of $\bar{\phi}$ would be $\Delta\bar{\phi} = \Delta S/(\frac{\partial S}{\partial\bar{\phi}})$,

$$(\Delta S)^2 = \langle\psi|(\hat{a}_1^\dagger\hat{a}_1)^2|\psi\rangle - \langle\psi|\hat{a}_1^\dagger\hat{a}_1|\psi\rangle^2. \tag{29}$$

In addition, in view of (28),

$$S=\langle\{0\}|\hat{\hat{a}}_1^\dagger\tilde{a}_1|\{0\}\rangle, \quad \Delta S=\langle\{0\}|(\hat{\hat{a}}_1^\dagger\hat{\hat{a}}_1)^2|\{0\}\rangle-\langle\{0\}|\hat{\hat{a}}_1^\dagger\hat{\hat{a}}_1|\{0\}\rangle^2. \tag{30}$$

A calculation shows that

$$\begin{aligned}\hat{\hat{a}}_1 &= \sum_{l\neq 1,j}(\cosh r U_{j1}^* U_{jl}e^{i\phi_j}\hat{a}_l - \sinh r U_{j1}U_{jl}^* e^{-i\phi_j}\hat{a}_l^\dagger)\\ &\quad+\sum_j |U_{j1}|^2[(\cosh^2 r e^{i\phi_j} - \sinh^2 r e^{-i\phi_j})\hat{a}_1\\ &\quad+2i\cosh r\sinh r\sin\phi_j\hat{a}_1^\dagger].\end{aligned} \tag{31}$$

Note the two different types of contribution to $\hat{\hat{a}}_1$ in Eq. (31): one much larger contribution from the mode which is initially squeezed (terms involving squares of the hyperbolic function) and the other smaller contribution from the initial unsqueezed modes. The signal $S$ is then given by

$$\begin{aligned}S &= \langle 0|\hat{\hat{a}}_1^\dagger\hat{\hat{a}}_1|0\rangle = I_1 + I_2,\\ I_1 &= 4\cosh^2 r\sinh^2 r\sum_{jk}|U_{ij}|^2|U_{1k}|^2\sin\phi_j\sin\theta_k,\\ I_2 &= \sinh^2 r\sum_{l\neq 1,jk} U_{j1}^* U_{k1}U_{k1}^* U_{j1}e^{-i(\phi_k-\phi_j)}.\end{aligned} \tag{32}$$

We simplify (32) using $U_{1j} = U_{j1}^* = \frac{1}{\sqrt{M}}$ and the unitary property of $\hat{U}$. We also assume that the $\phi$ are small so that we can expand exponentials in a power series. The first term in (32) is simple and equal to

$$I_1 \approx \bar{\phi}^2(4\cosh^2 r\sinh^2 r), \quad \bar{\phi} = \frac{\phi_1+\phi_2+\cdots+\phi_M}{M}. \tag{33}$$

The second term is more complicated:

$$I_2 \approx \Phi\sinh^2 r, \quad \Phi = \left(\frac{\phi_1^2+\phi_2^2+\cdots+\phi_M^2}{M} - \bar{\phi}^2\right). \tag{34}$$

Thus the signal at the output 1 is a function of two parameters of the network $\bar{\phi}$ and $\Phi$. For the simplicity of the measurement, we can assume large squeezing and drop the term $I_2$ in (34). Thus we can approximate the signal $S$ at the output port 1 as

$$S \approx I_1 = 4\cosh^2 r\sinh^2 r\,\bar{\phi}^2. \tag{35}$$

Note further that for small $\phi_j$, $\hat{\hat{a}}_1$ can be written as

$$\hat{\hat{a}}_1 = \hat{a}_1 - i\bar{\phi}(\cosh^2 r + \sinh^2 r)\hat{a}_1 + 2i\cosh r\sinh r\,\bar{\phi}\hat{a}_1^\dagger + R,$$

where $R$ is the contribution from $\hat{a}_l$ and $\hat{a}_l^\dagger$ with $l\neq 1$. The operator $R$ is important for the validity of the commutation relation $[\hat{\hat{a}}_1, \hat{\hat{a}}_1^\dagger] = 1$. It should be kept in mind that $R$ has terms only linear in hyperbolic functions $\cosh r$ and $\sinh r$. We compute $\Delta S$ to order $\bar{\phi}$:

$$\begin{aligned}(\Delta S)^2 &= \langle(\hat{\hat{a}}_1^\dagger\hat{\hat{a}}_1)^2\rangle - \langle\hat{\hat{a}}_1^\dagger\hat{\hat{a}}\rangle^2\\ &= \left\langle\hat{\hat{a}}_1^{\dagger 2}\hat{\hat{a}}_1^2\right\rangle + \langle\hat{\hat{a}}_1^\dagger\hat{\hat{a}}_1\rangle - \langle\hat{\hat{a}}_1^\dagger\hat{\hat{a}}_1\rangle^2\\ &\approx 2\langle\hat{\hat{a}}_1^\dagger\hat{\hat{a}}_1\rangle\\ &= 2S.\end{aligned} \tag{36}$$

On using (36) and (35), we obtain, for the error sensitivity,

$$\Delta\bar{\phi} = \Delta S\Big/\frac{\partial S}{\partial\bar{\phi}} = \frac{1}{2\sqrt{2}\sinh r\cosh r}, \tag{37}$$

which agrees with the QCRB derived in Appendix B. It should be recalled that the expression (37) is obtained for large $r$; otherwise $\Delta\bar{\phi}$ will be greater than (37).

We conclude this section by examining the case when the input is a single-mode squeezed coherent state, i.e.,

$$|\psi\rangle = \exp\left(\tfrac{1}{2}r\hat{a}^{\dagger 2} - \tfrac{1}{2}r\hat{a}^2\right)|\alpha\rangle,$$

where $|\alpha|^2$ is the average photon number in the coherent state. We assume that the coherent state has few photons. This was the case studied by Guo *et al.* [19]. The error sensitivity in the average phase of the network is obtained from measurements of the mean homodyne signal and fluctuations in the homodyne signal. Guo *et al.* [19] did the homodyne measurements on all the ports. We now present the results of the homodyne measurements at the output port 1 in the SU(1,1)-SU($m$) interferometer. For this purpose we use (31) as well as $\hat{a}_1|\alpha\rangle = \alpha|\alpha\rangle$ and $\hat{a}_l|0\rangle = 0$ ($l\neq 1$); hence

$$\begin{aligned}\langle\hat{\hat{a}}_1\rangle &= \alpha\left(\sum_j(2i\cosh r\sinh r\sin\phi_j - \sinh^2 r e^{-i\phi_j})|U_{j1}|^2\right)\\ &\cong \alpha + i\alpha\bar{\phi}e^{2r}\end{aligned} \tag{38}$$

for small $\phi$. We define the homodyne signal $S_{\text{HD}}$ as proportional to $\text{Im}\langle\hat{\hat{a}}_1\rangle$,

$$S_{\text{HD}} = \alpha\bar{\phi}e^{2r}, \quad \frac{\partial S_{\text{HD}}}{\partial\bar{\phi}} = \alpha e^{2r}. \tag{39}$$

The fluctuations in $S_{\text{HD}}$ for $\phi\to 0$ can be obtained by noting that $\hat{\hat{a}}_1 \to \hat{a}_1$ as $\phi\to 0$,

$$\Delta\hat{S}_{\text{HD}}^2 = \left\langle\left(\frac{\hat{a}_1-\hat{a}_1^\dagger}{2i}\right)^2\right\rangle - \left\langle\left(\frac{\hat{a}_1-\hat{a}_1^\dagger}{2i}\right)\right\rangle^2 = \frac{1}{4}, \tag{40}$$

and hence the error sensitivity is

$$(\Delta\bar{\phi})_{\rm HD} = \frac{e^{-2r}}{2\alpha}. \tag{41}$$

It is interesting that the error sensitivity measurement using only port 1 of the SU(1,1)-SU($m$) interferometer coincides with the one in the scheme of Guo *et al.* [19].

Before we conclude this section we like to relate (41) to the expression for the error sensitivity to the QFI using the input state as a squeezed coherent state. In this case the result (B12) of Appendix B is modified to

$$F_Q = 4\big[\langle\alpha|\hat{S}_1^\dagger(\hat{a}_1^\dagger\hat{a}_1)^2\hat{S}_1|\alpha\rangle - (\langle\alpha|\hat{S}_1^\dagger\hat{a}_1^\dagger\hat{a}_1\hat{S}_1|\alpha\rangle)^2\big], \tag{42}$$

which can be simplified to

$$\begin{aligned} F_Q = 4[&\sinh^4 r + \sinh^2 r + \sinh^2 r\cosh^2 r \\ &+ \alpha^2 e^{2r}(1 + 2\sinh^2 r + 2\sinh r\cosh r)]. \end{aligned} \tag{43}$$

Further simplification depends on the relative magnitudes of $|\alpha|^2$ and $\sinh^2 r$, i.e., photon numbers in the original coherent states and the number generated if initially $\alpha = 0$. Choosing $|\alpha|^2 \gg \sinh^2 r$ (as in experiments of Guo *et al.* [19]), $F_Q \approx 4\alpha^2 e^{4r}$, leading to the Cramér-Rao bound

$$(\Delta\phi)_{\rm QCRB} \approx \frac{e^{-2r}}{2\alpha} = (\Delta\bar{\phi})_{\rm HD}, \tag{44}$$

which coincides with the error sensitivity result (41).

## V. CONCLUSION

In this work we revealed the utility of SU(1,1)-SU($m$) interferometry in the context of sensing the parameters of a quantum network. We demonstrated this explicitly for the case of distributed phase sensing and force fields causing displacements. There is no need to find ways to measure the quantum Fisher information; the error sensitivity is sufficient to obtain Heisenberg sensitivity. This suggested alternative is simpler than homodyne measurements at many output ports. We add that a very [51] recent work discusses the use of a truncated SU(1,1) interferometer for distributed phase sensing. The SU(1,1)-SU($m$) interferometry is applicable to other parameters like a network of angular displacements.

## ACKNOWLEDGMENTS

The author is grateful for support through NSF Award No. 2426699, Air Force Office of Scientific Research Award No. FA-9550-20-1-0366, the Robert A. Welch Foundation Grant No. A-1943-20240404, and the Department of Energy, Fusion Energy Science, Award No. DE-SC0024882, IFE-STAR.

The author declares no conflicts of interest.

## DATA AVAILABILITY

No data were created or analyzed in this study.

## APPENDIX A: QCRB FOR THE AVERAGE DISPLACEMENT OF THE NETWORK

In this Appendix we present a detailed derivation of the quantum Fisher information and the QCRB for the averaged displacement $\bar{\alpha}$ of the network. The QFI can be calculated from the formal expression for $F_Q$,

$$F_Q = 4\,{\rm Re}(\langle\partial_{\bar{\alpha}}\psi|\partial_{\bar{\alpha}}\psi\rangle - |\langle\partial_{\bar{\alpha}}\psi|\psi\rangle|^2), \tag{A1}$$

where $|\psi\rangle$ is the state of $M$ modes just after the action of the network displacements,

$$|\psi\rangle = \prod_j D_j(\alpha_j)\hat{U}\hat{S}_1(\zeta)|\{0\}\rangle. \tag{A2}$$

We note that

$$\begin{aligned} |\partial_{\bar{\alpha}}\psi\rangle &= \frac{\partial}{\partial_{\bar{\alpha}}}\prod_j D_j(\alpha_j)\hat{U}\hat{S}_1(\zeta)|\{0\}\rangle \\ &= \frac{\partial}{\partial_{\bar{\alpha}}}\exp\left(\sum_j \alpha_j(\hat{a}_j^\dagger - \hat{a}_j)\right)\hat{U}\hat{S}_1(\zeta)|\{0\}\rangle \\ &= \frac{\partial}{\partial_{\bar{\alpha}}}\sum_j \alpha_j(\hat{a}_j^\dagger - \hat{a}_j)|\psi\rangle \\ &= \sum_j(\hat{a}_j^\dagger - \hat{a}_j)|\psi\rangle. \end{aligned} \tag{A3}$$

On substituting (A3) in (A1), we get

$$\begin{aligned} F_Q = 4\,{\rm Re}\Bigg[&-\langle\psi|\left(\sum_j(\hat{a}_j^\dagger - \hat{a}_j)\right)^2|\psi\rangle \\ &-|\langle\psi|\sum_j(\hat{a}_j^\dagger - \hat{a}_j)|\psi\rangle|^2\Bigg]. \end{aligned} \tag{A4}$$

We also note that

$$\begin{aligned} &\left(\prod_l D_l(\alpha_l)\right)^\dagger\left(\sum_j(\hat{a}_j^\dagger - \hat{a}_j)\right)\left(\prod_l D_\alpha(\alpha_l)\right) \\ &\quad= \sum_j[(\hat{a}_j^\dagger + \alpha_j) - (\hat{a}_j + \alpha_j)] = \sum_j(a_j^\dagger - \hat{a}_j), \end{aligned} \tag{A5}$$

as all the $\alpha$ are real. Thus Eq. (A4) can be written as

$$\begin{aligned} F_Q = 4\,{\rm Re}\Bigg[&-\langle\{0\}|\hat{S}_1^\dagger(\zeta)\hat{U}^\dagger\left(\sum_j(\hat{a}_j^\dagger - \hat{a}_j)\right)^2\hat{U}\hat{S}_1(\zeta)|\{0\}\rangle \\ &-|\langle\{0\}|\hat{S}_1^\dagger(\zeta)\hat{U}^\dagger\sum_j(\hat{a}_j^\dagger - \hat{a}_j)\hat{U}\hat{S}_1(\zeta)|\{0\}\rangle|^2\Bigg]. \end{aligned} \tag{A6}$$

The action of the $U$ on $\sum_j(\hat{a}_j^\dagger - \hat{a}_j)$ will produce a linear combination of the $\hat{a}_j$ and $\hat{a}_j^\dagger$ and the expectation values $\langle\{0\}|\hat{S}_1^\dagger(\zeta)\hat{a}_j\hat{S}_1(\zeta)|\{0\}\rangle$ are zero and hence the second term inside the large square brackets in (A6) vanishes. We now simplify the first term in (A6). For simplicity, we write $\hat{S}_1(\zeta)|\{0\}\rangle$ as $|\zeta\rangle$. The operator appearing in (A6) can be written in normal ordered form

$$\hat{U}^\dagger\left(\sum_{jl}\hat{a}_j^\dagger\hat{a}_l^\dagger + \sum_{jl}\hat{a}_j\hat{a}_l - 2\sum_{jl}\hat{a}_j^\dagger\hat{a}_l - M\right)\hat{U}, \tag{A7}$$

and then we use the fact that $\langle\zeta|\hat{a}_j^\dagger\hat{a}_l|\zeta\rangle \neq 0$ only if $j = l = 1$. We evaluate one of the terms in (A7) as a similar procedure would apply to the remaining terms:

$$\begin{aligned}\langle\zeta|\hat{U}^\dagger \sum_{jl}\hat{a}_j\hat{a}_l\hat{U}|\zeta\rangle &= \langle\zeta|\sum_{jl}\sum_{j'l'}U_{jj'}U_{ll'}\hat{a}_{j'}\hat{a}_{l'}|\zeta\rangle \\ &= \langle\zeta|\sum_{jl}U_{j1}U_{l1}\hat{a}_1\hat{a}_1|\zeta\rangle \\ &= \langle\zeta|\sum_{jl}\frac{1}{\sqrt{M}}\frac{1}{\sqrt{M}}\hat{a}_1^2|\zeta\rangle \\ &= M\langle\zeta|\hat{a}_1^2|\zeta\rangle. \end{aligned}\tag{A8}$$

Using similar calculations, the expression for $F_Q$ reduces to

$$\begin{aligned}F_Q &= 4M\,\mathrm{Re}(-\langle\zeta|\hat{a}_1^2|\zeta\rangle - \langle\zeta|\hat{a}_1^{\dagger 2}|\zeta\rangle + 2\langle\zeta|\hat{a}_1^\dagger\hat{a}_1|\zeta\rangle + 1) \\ &= 4M(1+2\sinh^2 r + 2\cosh r\sinh r) \\ &= 4Me^{2r}.\end{aligned}\tag{A9}$$

Hence the QCRB for $\bar{\alpha}$ is

$$(\Delta\bar{\alpha})_{\rm QCRB} = \frac{1}{\sqrt{F_Q}} = \frac{1}{2\sqrt{M}e^r}.\tag{A10}$$

## APPENDIX B: QUANTUM FISHER INFORMATION FOR THE AVERAGE PHASE OF THE DISTRIBUTED NETWORK

The state under consideration after transformations by the unitary operators $U$, $S_1$, and $R(\phi_i)$ is

$$|\psi\rangle = \prod_j e^{i\phi_j\hat{a}_j^\dagger\hat{a}_j}\hat{U}\hat{S}_1|\{0\}\rangle.\tag{B1}$$

The average phase is

$$\bar{\phi} = \frac{\phi_1+\phi_2+\cdots+\phi_M}{M}.\tag{B2}$$

The QFI will be given by

$$F_Q = 4(\langle\partial_{\bar{\phi}}\psi|\partial_{\bar{\phi}}\psi\rangle - \langle\partial_{\bar{\phi}}\psi|\psi\rangle\langle\psi|\partial_{\bar{\phi}}\psi\rangle)\tag{B3}$$

in the limit $\phi_i \to 0$. In order to calculate (B3), we need to transform the $\phi_i$ to a new set of variables $\eta_1, \ldots, \eta_M$ such that $\eta_1 = \sqrt{M}\bar{\phi}$, where $s$ is a number and

$$\eta_i = \sum_j V_{ij}\phi_j, \quad V_{1j} = \frac{1}{\sqrt{M}},\tag{B4}$$

where $V$ is an orthogonal matrix so that

$$\phi_i = \sum_j (V^{-1})_{ij}\eta_j\tag{B5}$$

$$= \sum_j V_{ji}\eta_j = V_{1i}\eta_1 + \sum_{j\neq i}V_{ji}\eta_j.\tag{B6}$$

As an explicit example of $V$ of $V = 4$ modes will be

$$V = \frac{1}{2}\begin{pmatrix}1 & 1 & 1 & 1\\ 1 & -1 & 1 & -1\\ 1 & 1 & -1 & -1\\ 1 & -1 & -1 & 1\end{pmatrix}.\tag{B7}$$

Therefore, the rotation operator in (B1) becomes

$$\begin{aligned}&\prod_j \exp(i\phi_j\hat{a}_j^\dagger\hat{a}_j) \\ &\quad = \exp\left(i\bar{\phi}\sum_i\sqrt{M}\,V_{1i}\hat{a}_i^\dagger\hat{a}_i + i\sum_{i\neq j}V_{ji}\eta_j\hat{a}_i^\dagger\hat{a}_i\right).\end{aligned}\tag{B8}$$

It is important to note that all the operators appearing in the exponent commute with each other, i.e., $[\hat{a}_i^\dagger\hat{a}_i, \hat{a}_j^\dagger\hat{a}_j] = 0\,\forall\, i, j$. Thus the derivative with respect to $\eta_1$ will yield

$$(\partial_{\bar{\phi}}|\psi\rangle)_{\phi_i\to 0} = \left(i\sqrt{M}\sum_i V_{1i}\hat{a}_i^\dagger\hat{a}_i\right)\hat{U}\hat{S}_1|\{0\}\rangle,\tag{B9}$$

and hence (B3) reduces to

$$\begin{aligned}F_Q &= 4\left[\langle\{0\}|\hat{S}_1^\dagger\hat{U}^\dagger\left(\sum_j\hat{a}_j^\dagger\hat{a}_j\right)^2\hat{U}\hat{S}_1|\{0\}\rangle \right.\\ &\qquad \left. - \left(\langle\{0\}|\hat{S}_1^\dagger\hat{U}^\dagger\sum_j\hat{a}_j^\dagger\hat{a}_j\hat{U}\hat{S}_1|\{0\}\rangle\right)^2\right] \\ &= 4\left[\langle\{0\}|\hat{S}_1^\dagger\left(\sum_j\hat{a}_j^\dagger\hat{a}_j\right)^2\hat{S}_1|\{0\}\rangle \right.\\ &\qquad \left. - \left(\langle\{0\}|\hat{S}_1^\dagger\sum_j\hat{a}_j^\dagger\hat{a}_j\hat{S}_1|\{0\}\rangle\right)^2\right].\end{aligned}\tag{B10}$$

Since $\sum_i\hat{a}_i^\dagger\hat{a}_i$ is conserved under unitary transformation, noting that $\hat{S}_1$ acts only on mode 1 and

$$\hat{a}_j^\dagger\hat{a}_j\hat{S}_1|\{0\}\rangle = 0\,\forall\, j\neq 1,\tag{B11}$$

Eq. (B10) reduces to

$$F_Q = 4\langle 0|\hat{S}_1^\dagger(\hat{a}_1^\dagger\hat{a}_1)^2\hat{S}_1|0\rangle - 4(\langle 0|\hat{S}_1^\dagger\hat{a}_1^\dagger\hat{a}_1\hat{S}_1|0\rangle)^2.\tag{B12}$$

The expression (B12) is the well-known QFI for phase using squeezed vacuum,

$$F_Q = 8(\sinh^2 r)(1+\sinh^2 r),\tag{B13}$$

and hence

$$(\Delta\bar{\phi})_{\rm QCRB} = \frac{1}{\sqrt{F_Q}} = \frac{1}{2\sqrt{2}\sinh r\cosh r},\tag{B14}$$

which is the Heisenberg limit for the average phase $\bar{\phi}$ as the total number of input photons in the network is $\sinh^2(r)$, which is distributed in $M$ nodes.

## APPENDIX C: EFFECT OF THE DETECTOR EFFICIENCY

In this Appendix we briefly examine the effect of the detector efficiency given by $\eta$. The input flux $\langle a^\dagger a\rangle$ is reduced to $\eta\langle a^\dagger a\rangle$. For simplicity, we restrict our analysis to the distributed displacements. In this case it was shown after (21) that the state of the field at output port 1 is a coherent state $\bar{\alpha}\sqrt{M}e^r$. It is known that the detector efficiency factor $\eta$ will transform the coherent state into a coherent state with reduced amplitude $\bar{\alpha}\sqrt{M\eta}e^r$. The error sensitivity can be recalculated following the method outlined before (23) with the result

$$\Delta\bar{\alpha} = \frac{1}{2\sqrt{\eta M}e^r}. \tag{C1}$$

Thus the detector efficiency worsens the error sensitivity by the factor $\frac{1}{\sqrt{\eta}}$.